\documentclass[journal]{IEEEtran}
\date{}
\usepackage{lineno}
\usepackage{graphicx}
\usepackage{cite}
\usepackage{amsmath}
\usepackage[usenames,dvipsnames]{color}
\usepackage{soul}
\setstcolor{red}
\setul{}{1.5pt}

\usepackage[labelfont=bf]{caption}
\usepackage[strict]{changepage}

\begin{document}
\title{Magneto-Mechanical Transmitters for Ultra-Low Frequency Near-field Communication}

\author{Rhinithaa~P.~Thanalakshme,
        Ali~Kanj,
        JunHwan~Kim,
        Elias~Wilken-Resman,
        Jiheng~Jing,
        Inbar~H.~Grinberg,
        Jennifer~T.~Bernhard,
        Sameh~Tawfick,
        ~and~Gaurav~Bahl
\thanks{R. P. Thanalakshme, A. Kanj, J. Kim, J. Jing, I. H. Grinberg, S. Tawfick, and G. Bahl are affiliated with the Department of Mechanical Science and Engineering, University of Illinois at Urbana-Champaign, Urbana, IL 61801 USA.}%
\thanks{E. Wilken-Resman and J. T. Bernhard are affiliated with the Department of Electrical and Computer Engineering, University of Illinois at Urbana-Champaign, Urbana, IL 61801 USA.}}

\maketitle

\begin{abstract}
Electromagnetic signals in the ultra-low frequency (ULF) range below 3 kHz are well suited for underwater and underground wireless communication thanks to low signal attenuation and high penetration depth. However, it is challenging to design ULF transmitters that are simultaneously compact and energy efficient using traditional approaches, e.g., using coils or dipole antennas. Recent works have considered magneto-mechanical alternatives, in which ULF magnetic fields are generated using the motion of permanent magnets, since they enable extremely compact ULF transmitters that can operate with low energy consumption and are suitable for human-portable applications. Here we explore the design and operating principles of resonant magneto-mechanical transmitters (MMT) that operate over frequencies spanning a few 10's of Hz up to 1 kHz. We experimentally demonstrate two types of MMT designs using both single-rotor and multi-rotor architectures. We study the nonlinear electro-mechanical dynamics of MMTs using point dipole approximation and magneto-static simulations. We further experimentally explore techniques to control the operation frequency and demonstrate amplitude modulation up to 10 bits-per-second. \end{abstract}


\definecolor{limegreen}{rgb}{0.2, 0.8, 0.2}
\definecolor{forestgreen}{rgb}{0.13, 0.55, 0.13}
\definecolor{greenhtml}{rgb}{0.0, 0.5, 0.0}

\section{Introduction}

Oscillatory sources of electromagnetic waves play a vital role in communications, navigation, and sensing systems \cite{RFCh1, RFCh2, RFCh3, EMwave,hfreq1,hfreq2}. While radio-frequency signals in the MHz-GHz range provide this functionality for above-ground communications, they experience significant attenuation when traveling through conductive media like seawater, rocks, and soil \cite{skineff}. The solution for these situations is to instead use ultra-low frequency (ULF) carrier signals below 3 kHz as they provide orders-of-magnitude greater signal penetration depth and lower signal attenuation than radio frequencies. Short range ULF communication systems, in particular, can rely on a non-propagating or quasi-static approach using the magnetic field generated by an oscillating magnetic dipole as the carrier signal, e.g., in pipe monitoring \cite{app1}, seismic sensors \cite{app2}, or underwater vehicles \cite{app3}. A straightforward inductive coupling implementation is often invoked for this \cite{mi1,mi2,mi3}, but needs rather large (meter scale) coils operating as transmitter-receiver pairs. While it is desirable to reduce the coil size to the centimeter scale to facilitate mobility, this is technically very challenging due to the heat generation from large currents in small coils \cite{EMwave} and due to parasitic capacitance issues that impact power transfer \cite{resmi}.

Recently, an alternative approach has been considered to produce modulated ULF magnetic fields, by leveraging the rotational or oscillatory motion of permanent magnets due to their very high residual flux density \cite{MA1,MA2,MA3, MA4,MA5,MA6,UCLA}. These systems often resemble an electric motor in their operating principles, and use single-degree-of-freedom magnetized rotors that are brought into motion using one or more drive coils. 

The most attractive feature of this method is the zero power requirement for the generation of the magnetic field. Instead, energy is now only consumed in the physical motion required for producing the harmonic carrier signal, and for the secondary modulation applied for transmitting data, thereby transforming the engineering challenges to different domains. When the carrier signal is generated using full rotation \cite{MA2,MA3,MA4,MA5,MA6}, data can be encoded by varying the rotational speed, resulting in a frequency modulation (FM) method. This modulation typically consumes a significant amount of energy per bit since a slow down/speed up operation must be used for data transmission, although regenerative braking methods may be considered. Alternatively, mu-metal shielding and switching circuits can also be invoked \cite{MA_mod1} for achieving amplitude modulation while the rotor maintains a constant rotational speed. 

 In this work, we describe a magneto-mechanical transmitter (MMT) design that utilizes resonant angular oscillatory motion of a single or multiple permanent magnet rotors to produce the ULF carrier signal (a variant of this design has been explored in other recent works \cite{MA6,UCLA}). As we show, data can be encoded on the carrier using an amplitude modulation approach. We discuss how the carrier signal frequency is set by the mechanical resonance of the synchronized torsional mode of the rotors, while the carrier amplitude is determined by both the sum of total magnetic dipole moment of the rotors as well as the amplitude of rotational motion. This resonant approach in particular enables the generation of large carrier amplitude for small electrical input provided to the drive coil. We experimentally demonstrate our approach using two MMT systems; one that uses a single-rotor design for operation below 200 Hz, and a multi-rotor design for operation up to 1 kHz. Additionally, we demonstrate how the use of tunable magnetic springs (i.e., magnetic stators) allows tuning of the carrier frequency, and discuss how we can achieve uniform motional distribution across the individual rotors of a multi-rotor MMT. 

\section{Design and analysis}

\subsection{Working principle}

The two principal components of an MMT are a torsional mechanical resonator and an adjacent driving coil. The mechanical resonator can be composed of one or more proof masses having permanent magnetic dipole moment. These permanent magnets are suspended with a single-axis rotational degree of freedom, and are provided torsional stiffness using magnetic restoring torque. \cite{magspring1,magspring2,magspring3}. A time-harmonic electrical signal provided to the coil produces a time-harmonic torque on this resonator, which in turn produces the effect in the near field of an oscillating magnetic field. The frequency of the magnetic signal is therefore set by the externally provided electrical input, but is maximized near the mechanical resonance. At any point in space, the amplitude of the time-varying component of the magnetic field is proportional to the total magnetic moment of the oscillating magnets, and is further scaled by the amplitude of the torsional motion. It is this time-varying component of the magnetic field that serves as the carrier signal for this ULF transmitter design.

\begin{figure}[t]
    \centering
	\includegraphics[width=0.75\linewidth]{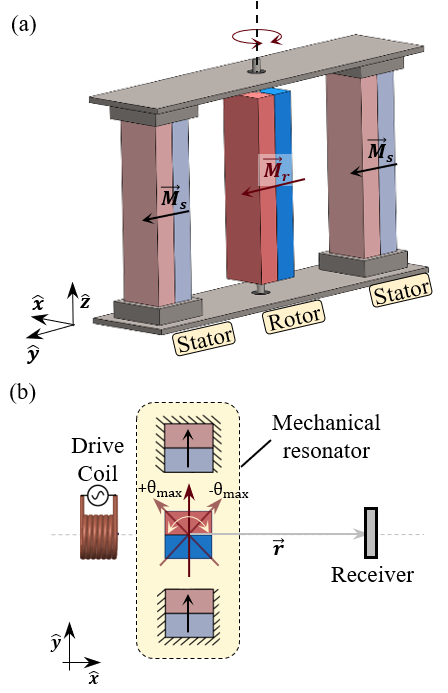}
    	\caption{Schematic of a single-rotor magneto-mechanical transmitter (MMT) 
    	\textbf{(a)} The mechanical resonator component of the MMT is comprised of a bearing-mounted permanent magnet with magnetic moment $\vec{M}_r$ that is free to oscillate about $\hat{z}$. Torsional stiffness is provided by two fixed magnets each with magnetic moment $\vec{M}_s$. 
    	\textbf{(b)} Top view of the complete MMT. The drive coil supplied with an ac voltage is placed adjacent to the rotor such that its axis is directed along $\hat{x}$. The rotor undergoes oscillations with amplitude of $\pm\theta_\textrm{max}$ and the receiver location with respect to the rotor's center is denoted by $\vec{r}$.
    	} 
    	\label{device}
\end{figure}

A schematic of the mechanical resonator is presented in Fig.~\ref{device}a. The magnetic proof mass (or rotor) is a permanent magnet that is mounted on two rotational bearings; these bearings provide the rotor the freedom to rotate about an axis orthogonal to its magnetic dipole. We place fixed magnets (called stators) adjacent to this rotor, which provide the restoring magnetic torque. The stators are assembled such that their dipoles are oriented along the same direction, which causes the rotor magnet (magnetic moment $\vec{M}_r$) to self-align with the stator magnetic field. Any angular displacement of the rotor from this equilibrium results in a restoring torque due to the magnetic interaction with the stator magnetic field giving rise to an effective torsional stiffness. 

While stators are not strictly essential in a multi-rotor MMT \cite{UCLA}, we will show later that they are very beneficial for designing a more uniform and reproducible synchronized mode for the rotors. 

All the mechanical losses in the resonator can be modeled by a lumped viscous damping coefficient $b$, implying that the assembly in Fig.~\ref{device}a is effectively a resonant mass-spring-damper system. The drive coil is placed adjacent to the resonator assembly such that a current through the coil induces a magnetic field that is orthogonal to the equilibrium position of the rotor's magnetic dipole; this induced magnetic field is capable of applying a torque on the rotor in order to drive its oscillations (Fig.~\ref{device}b). The coil is supplied with an ac voltage at a desired frequency, which results in oscillation of the rotor, and the accompanying time-varying magnetic field at the intended receiver.

\subsection{Carrier magnetic field generated by the MMT}

In order to estimate the carrier magnetic field produced by the MMT, we can approximate the rotor as a point dipole with moment $\Vec{M}_r$ initially oriented along $\hat{y}$ as shown in Fig.~\ref{device}b. At rest, the rotor only produces a dc magnetic field in its vicinity. When the rotor undergoes time-harmonic angular displacement $\theta = \theta_\textrm{max} \sin (\omega t)$ about $\hat{z}$, the magnetic field at any receiver position $\Vec{r}$ now exhibits an additional time dependence (implicit in $\theta$), and is evaluated by
\begin{equation}\label{bfieldeq}
    {\vec{B}}(\vec{r},\theta) = \frac{\mu_o}{4\pi|\vec{r}|^3} \Big[3 \hat{r} \left({\vec{M}}_r(\theta) \cdot \hat{r}\right)-{\vec{M}}_r (\theta) \Big]
\end{equation}
where $\mu_o=4\pi \times 10^{-7}$ N/A$^2$ is the permeability of free space. In this work, we are interested in the amplitude of the sinusoidal $\omega$-frequency component of magnetic field at the receiver. For the configuration shown in Fig.~\ref{device}b, we find that this component of ${\vec{B}}$ is maximized in the $\hat{x}$ direction. As a result, a vector receiver placed at $\vec{r} = r_x \hat{x}$ and oriented to capture the $\hat{x}$-directed field measures
\begin{equation}\label{bamp_t}
    {B}_\textrm{mech}(r_x,t) = -\frac{\mu_o M_r \sin(\theta)} {2 \, \pi r_x^3}
\end{equation}
as a function of instantaneous angular displacement $\theta$. Here, we use the `mech' subscript to indicate that the field is produced from mechanical motion of the permanent dipole. The rms value of ${B}_\textrm{mech}(r_x,t)$ can then be evaluated as
\begin{equation}\label{bamp}
    {B}_\textrm{mech,rms}(r_x) = \frac{\mu_o M_r \sin(\theta_\textrm{max})} {2\sqrt{2} \, \pi r_x^3}~.
\end{equation}
Eq.~\ref{bamp} serves as a design guide for rotor size and oscillation amplitude needed to achieve a desired field at the receiver. Since portable high-sensitivity magnetometers \cite{magnetometer1,magnetometer2} can detect femtotesla-level magnetic fields, we set the desired rms amplitude at a receiver distance of 1 km as 1 fT. For the single-rotor design we explore in this paper, we use commercially available cuboidal N52 grade NdFeB magnets having residual flux density 1.48~kGa for both the rotors and the stators, with axial length 50.8 mm and identical square cross-section of side length 12.7 mm. Eq.~\ref{bamp} then indicates that $\pm45^\circ$ oscillation is required to achieve the desired magnetic field.

\subsection{Resonance of the mechanical resonator}
\label{sec:resonance_theory}

Since the angular mechanical motion of the rotor is maximized near resonance, the largest oscillating magnetic field contributed by the mechanics is produced at the same frequency. To determine this frequency, we extend the point dipole model presented in the previous section by including the restoring effect of stator magnets on the rotor.

When a rotor with dipole moment $\vec{M}_r$ (Fig.~\ref{device}a) and mass moment of inertia $I$ is placed in the stator magnetic field, it experiences a nonlinear restoring torque ${\tau}_\textrm{rs}$ \cite{inbar} along $-\hat{z}$ (up to the third order term) from two stators given by 
\begin{equation} \label{stiffness}
    {\tau}_\textrm{rs} = \frac{\mu_o M_r M_s}{2 \pi {d}_\textrm{rs}^3}\left(2 \theta-\frac{\theta^3}{3}\right) = \kappa_1\theta + \kappa_3\theta^3
\end{equation}
\noindent
where $M_s$ represents the magnitude of dipole moment of a single stator and $d_\textrm{rs}$ represents the center-to-center distance between the rotor and stator magnets. We find that the nonlinear stiffness term $\kappa_3 < 0$, implying that the oscillating dipole is expected to show a softening nonlinearity \cite{inbar}. For small oscillation amplitudes, the torsional stiffness contributed by both stators can then be evaluated using the linear stiffness term as $\kappa_1 \approx {\mu_o M_r M_s}/{\pi d_\textrm{rs}^3}$ and the natural frequency of the mechanical resonator is then obtained as $\omega_0 \approx \sqrt{\kappa_1/I}$. When the oscillation amplitude is large the nonlinear term becomes more relevant, and we can predict the mechanical resonance frequency of the nonlinear resonator through the backbone curve \cite{nayfehnonlinear} 
\begin{equation}\label{backbone}
    \omega_\textrm{nonlin} = \omega_o + \frac{3}{8}\frac{\kappa_3}{\sqrt{\kappa_1I}}\theta_\textrm{max}^2
\end{equation}
where $\omega_o$ is the natural frequency at low amplitudes where the nonlinearity can be ignored.

\begin{figure}[!t]
     
    \centering
    \includegraphics[width= \linewidth]{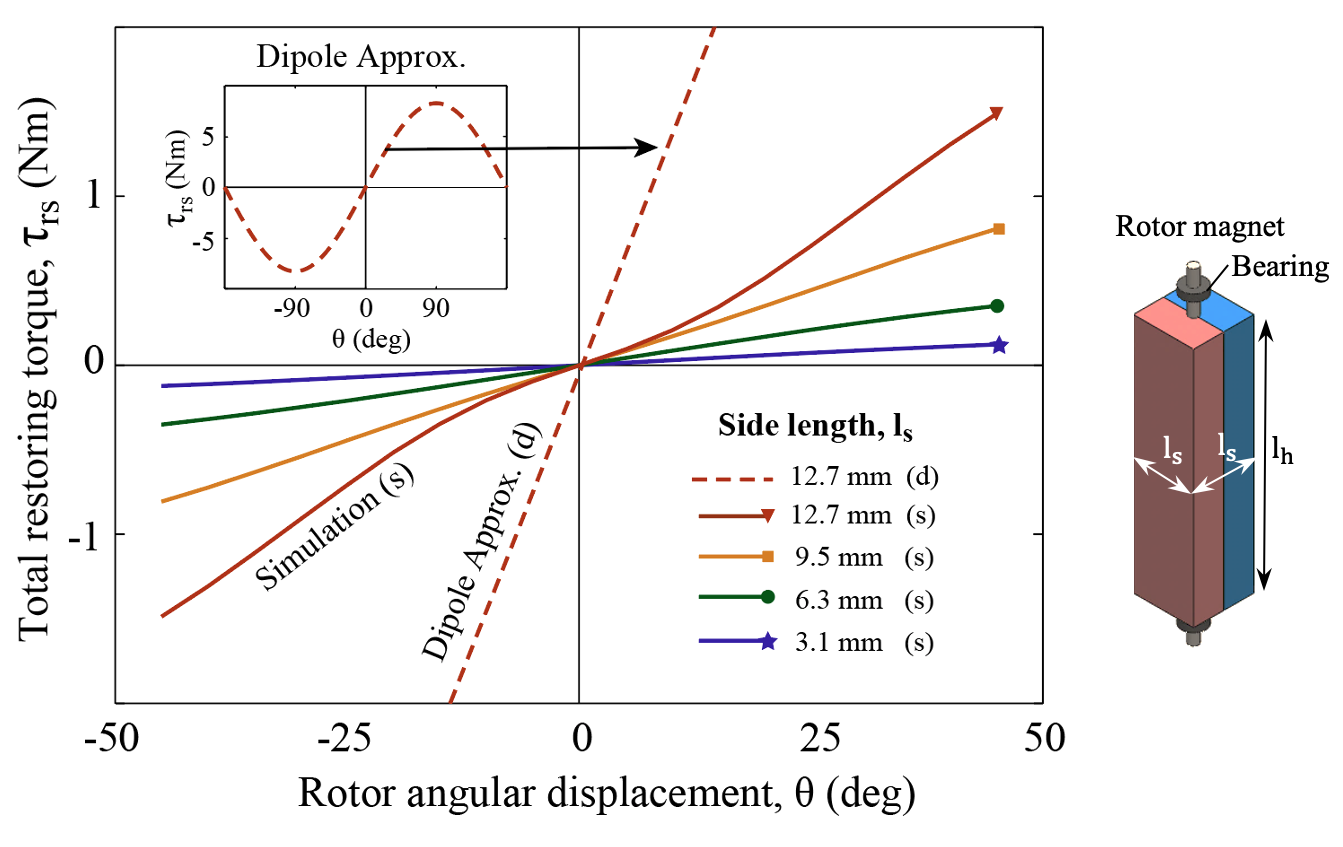}
    \caption{ Dependence of the magnetic restoring torque on rotor angular displacement for $\theta$ between $\pm45^\circ$. Estimates for the restoring torque on the rotor due to the two stators are computed using the dipole approximation (denoted by d), and the numerical solutions from ANSYS Maxwell magnetostatic simulations (denoted by s). The rotor and stators are N52 NdFeB cuboidal magnets with residual flux density 1.48 kGa and axial length $l_h = \textrm{50.8 mm}$. In the four simulations, the rotor side length $l_s$ is varied between 12.7 mm and 3.1 mm while fixing the stator side length to 12.7 mm and rotor-stator distance to 16.5 mm. Inset shows the dipole model solution for the case with rotor $l_s$ = 12.7 mm. }\label{torquefig}
      
\end{figure}
 
This point dipole model, however, does not fully capture the local non-uniformity of the magnetic field due to the geometry of the magnets \cite{nonlin1,nonlin2,nonlin3}. In our case, the geometrical effect is quite significant as the rotor-stator distance is comparable to the physical dimensions of the magnets themselves. A better estimate for the torsional magnetic stiffness can instead be obtained by numerically computing the $\tau_\textrm{rs}$ vs. $\theta$ curve via finite element simulations. Here, we use ANSYS Maxwell to model the torsional stiffness. Fig.~\ref{torquefig} shows example $\tau_\textrm{rs}$ vs. $\theta$ curves for a variety of rotor cross-sections, for angular displacements spanning $\pm45^\circ$. For the example case of a square rotor of side length $l_s = 12.7$ mm, we see that the dipole approximation overestimates the torsional stiffness compared to the finite element simulation. Notably, the softening effect is only apparent in the dipole model for extremely large angular displacement approaching $\pm 90^\circ$. However, the corresponding numerical simulation shows a stiffening nonlinearity even for oscillations spanning $\pm 15^\circ$, which is a result of the near-field effect captured by the finite element model. We now use the numerical results to fit the linear and nonlinear stiffness coefficients in Eq.~\ref{stiffness}, and then use Eq.~\ref{backbone} to estimate a mechanical resonance frequency of 168 Hz for the $l_s = 12.7$ mm device described in Fig.~\ref{torquefig} oscillating up to $\theta_\textrm{max}=45^\circ$. We provide corresponding experimental validation for this configuration in section \ref{results_singlerotor_freq}. 

To further explore the effect of magnet geometry on the restoring torque, we simulated a range of resonator designs by varying the side length of the square cross-section rotor. Alongside, we keep the rotor axial length $l_h$, stator magnet dimensions, and rotor-stator distance unchanged. As the rotor cross-section is made smaller, the near-field effect is less pronounced, as a result of which the stiffening nonlinearity vanishes. Consequently, the geometry of the magnet determines the nature of nonlinear stiffness in the resonator. This helps understand the difference in the stiffness trends shown by the single-rotor and multi-rotor MMT designs presented later.

\subsection{Equivalent circuit model}\label{se:LEM}

In order to understand the combined electro-mechanical response and power consumption for an MMT, we develop an equivalent circuit model that includes both electrical and mechanical components. This type of lumped element model is extensively used in analyzing systems with coupled electrical and mechanical domains \cite{LEM1,LEM2,LEM3,LEM4}. 

\begin{figure}
     
    \centering
    \includegraphics[width=0.8\linewidth]{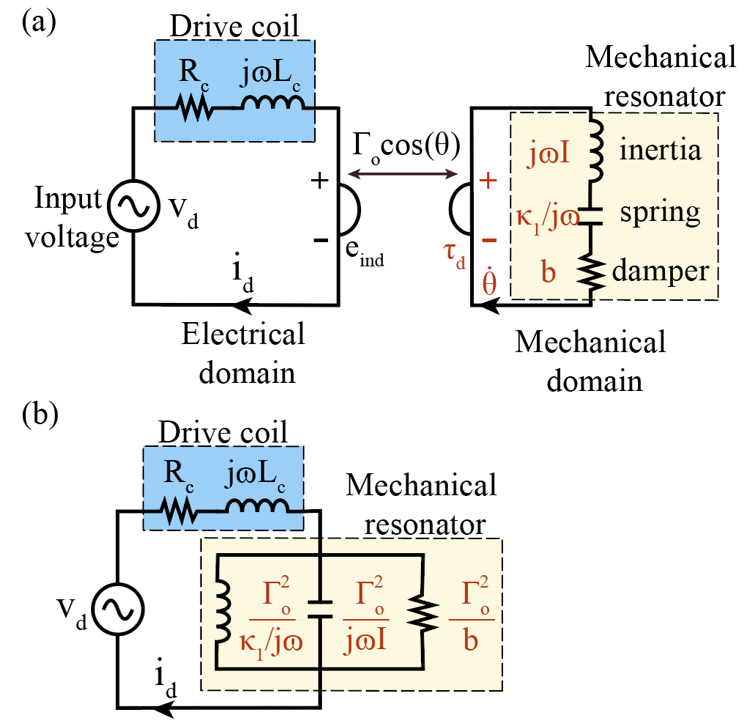}
    \caption{  Equivalent circuit of MMT using lumped elements. \textbf{(a)} A gyrator is used to couple the drive coil in the electrical domain with the resonator in the mechanical domain via instantaneous coupling coefficient $\Gamma_o \cos(\theta)$. \textbf{(b)} Electrical domain representation of the MMT obtained by transforming the series RLC configuration of the mechanical resonator using the linear coupling coefficient $\Gamma_\textrm{o}$ under small angle approximation.
}\label{circuit}
      
\end{figure}

By convention \cite{LEMgyrator, F-Vanology, LEM1,LEM4}, quantities that are measured across a lumped element are called ‘effort’, and those passing through are called ‘flow’. In Fig.~\ref{circuit}a, the circuit on the left represents the equivalent lumped element model of the drive coil and the circuit on the right represents the mechanical resonator. The drive coil can be modeled as a series RL circuit, in which the flow variable is the current flow through the coil and the effort variable is the voltage across the elements. The mechanical resonator can be modeled as a series RLC circuit, in which the flow variable is the angular velocity of the rotor and the effort variable is the torque acting on it \cite{LEMgyrator}. In this mechanical-equivalent circuit, the resistor represents the mechanical damping $b$, the inductor represents the rotor moment of inertia $I$, and the capacitor represents the linearized torsional stiffness $\kappa_1$ of the mechanical resonator. When the drive coil is active, the drive current exerts a torque on the mechanical resonator. Similarly, the angular velocity of the mechanical resonator induces a back emf in the drive coil. Due to this relationship where the flow variable in one circuit maps to the effort variable in the other circuit, we use a two-port gyrator to represent the coupling between these two equivalent circuits \cite{LEMgyrator}, as shown in Fig.~\ref{circuit}a. 
 
\begin{figure*}[ht]
\centering
	\includegraphics[width=0.75\textwidth]{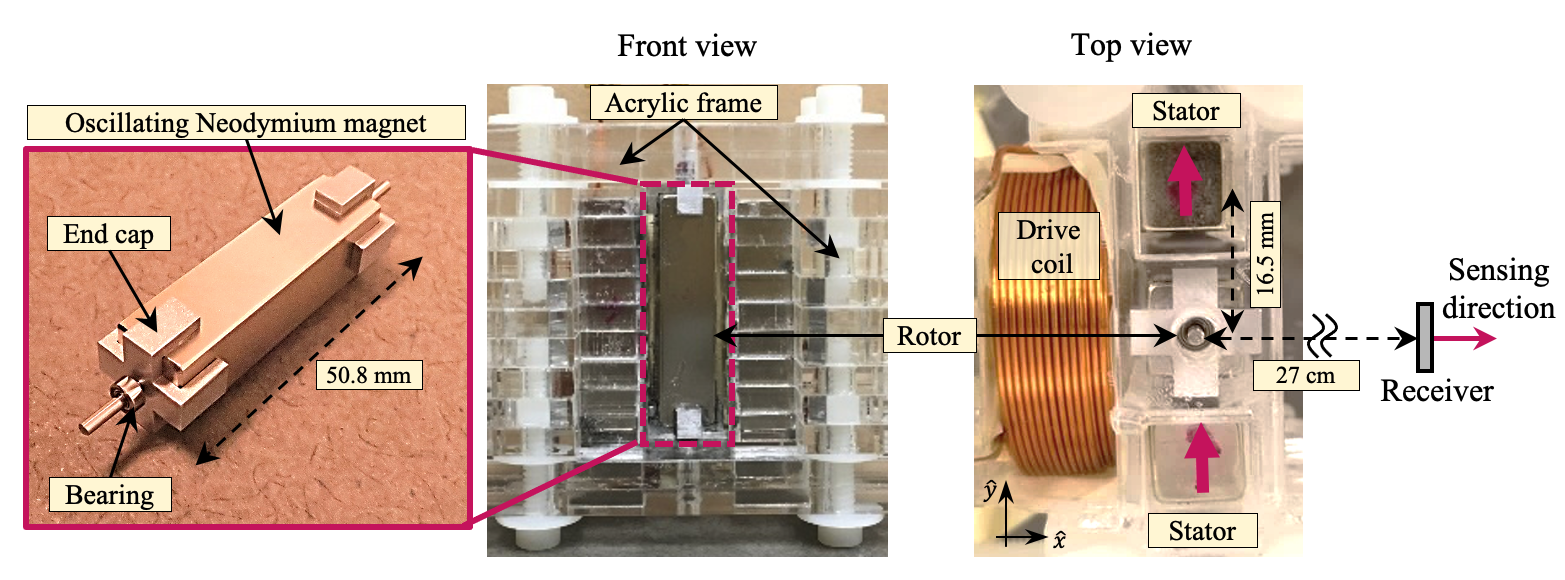}
    \caption{Construction of single-rotor MMT using cuboidal rotor and stators, along with the drive coil and flux-gate magnetometer as receiver.}
    \label{prototype}
\end{figure*}

We can now develop the coupling equations between the mechanical and the electrical elements to determine the coupling coefficient described by the gyrator. The drive coil with $N_c$ turns and cross-section area $A_c$ is supplied with a drive voltage $v_\textrm{d}(t)=\sqrt{2} \, v_\textrm{d,rms} \, \textrm{sin}(\omega t)$, resulting in a sinusoidal current $i_d$ flowing through the coil. The adopted sign convention for positive $i_d$ produces a magnetic field along $\hat{x}$ (Fig.~\ref{device}b) that exerts an instantaneous torque $\vec{\tau}_d$ on the rotor about $\hat{z}$. The magnetic field generated by the coil on the rotor at distance $d_{cr}$ is given by
\begin{linenomath}
\begin{align}\label{bfieldc}
    {\vec{B}_\textrm{c}(d_{cr},t)} = \frac{\mu_o N_c A_c}{2\pi d_{cr}^3} \, i_d \, \hat{x}
\end{align}
\end{linenomath}
and the instantaneous torque is given by
\begin{linenomath}
\begin{align}    
   \nonumber {\tau_d}= (\vec{M_r} \times {\vec{B}_\textrm{c}}) \cdot \hat{z}  &= -\frac{\mu_o M_r N_c A_c \cos (\theta) }{2\pi d_\textrm{cr}^3} \, i_d 
    \\&= -\Gamma_o \cos(\theta) \, i_d 
\end{align}  
\end{linenomath}
for a rotor deflected instantaneously to rotational angle $\theta$. If the rotor is in motion with angular velocity $\dot{\theta}$, the magnetic flux $\Phi$ threading through the coil will be a time-varying quantity and will induce a back emf $e_\textrm{ind}$ across the coil terminals, as follows
\begin{linenomath}
\begin{align}
    e_\textrm{ind}=-N_c\frac{d\Phi}{dt}=\frac{\mu_o M_r N_c A_c \cos (\theta) }{2\pi d_\textrm{cr}^3} \, \dot{\theta} = \Gamma_o \cos(\theta) \, \dot{\theta} ~.
\end{align}
\end{linenomath}
To facilitate a linear analysis we can assume that the angle of displacement is small i.e., $\cos (\theta)\approx 1$. With this we are able to introduce the gyrator coupling coefficient $\Gamma_o$ into the equivalent circuit model. This approximation ultimately helps develop an understanding of the electrical behavior and guides the design of the MMT.

The frequency response of the MMT can be studied in the electrical domain by deriving an expression for the total impedance of the MMT from the perspective of the drive voltage source $v_d$. For this purpose, the mechanical elements are transferred to the electrical domain using the impedance transformation \cite{LEM1}, as illustrated in Fig.~\ref{circuit}b. The total impedance $Z_\text{tot}$ (in frequency domain) measured across the voltage source is then given by
\begin{align}\label{Z}
  \nonumber  Z_\textrm{tot} = \left[R_c+\frac{\Gamma_\textrm{o}^2 \omega^2 b}{(\kappa_1-\omega^2I)^2+(\omega b)^2}\right] 
            \\+j \left[\omega L_c+\frac{\Gamma_\textrm{o}^2 \omega (\kappa_1-\omega^2I)}{(\kappa_1-\omega^2I)^2+(\omega b)^2}\right] ~.
\end{align}
 $Z_\textrm{tot}$ exhibits a resonant peak that is slightly offset from $\omega_o=\sqrt{\kappa_1/I}$ due to the drive coil reactance. Since we are interested in maximizing the mechanically-produced magnetic field at the receiver, we instead compute the frequency at which the angular oscillation amplitude of the rotor is maximized. The frequency domain voltage-to-angular-displacement transfer function can be expressed as
\begin{equation}\label{thetaTF}
    \frac{\Theta}{V_d} = \frac{1}{j\omega\Gamma_\textrm{o}}\frac{Z_\textrm{{mech}}}{Z_\textrm{tot}}
\end{equation}
\noindent
where $\Theta$ and $V_d$ represent the angular displacement and voltage input in the frequency domain, respectively, and $Z_\textrm{mech} = {\Gamma_o^2}/({\frac{\kappa_1}{j\omega}+ b + j{\omega}I})$ is the effective impedance due to the mechanical resonator. While the complete closed-form expression of the resulting mechanical resonance frequency is rather complex and provides little insight, we can simplify it in the case of small damping ($R_c = 0 \; \textrm{and}\; b=0$) as follows
\begin{equation}\label{mech_freq}
    \omega_\textrm{mech} = \sqrt{ \frac{\kappa_1  \, + \,  \Gamma_o^2/L_c }{I}} ~.
\end{equation}
This result indicates that the mechanical resonance frequency of an MMT (in the low loss regime) is greater than its independent resonance $\omega_o$, and depends both on the coupling coefficient and on the drive coil inductance. For a typical single-rotor MMT considered in this work ($\omega_o \approx 168$ Hz, $\Gamma_o \approx 10^{-2}$~Nm/A, and $I \approx 1.6 \times 10^{-6}$ kg m${}^2$), the additional contribution ${\Gamma_o^2}/{L_c I}$ is much smaller than the mechanical-only contribution ${\kappa_1}/{I}$. We can therefore approximate $\omega_\textrm{mech} \approx \omega_o$ with the remaining offset adjusted experimentally. Even if we do not assume the damping to be negligible, (typically the mechanical damping is $b \approx 10^{-5}-10^{-4}$ N m s / rad and $R_c \approx 0.1-2$ ohms), we find that the shift in mechanical resonance frequency is less than 5$\%$ of $\omega_o$. Hence, for practical purposes, $\omega_o$ can be used for designing the MMT with reasonable accuracy.

Another parameter of interest is the average power consumed by the MMT. Using the circuit model in Fig.~\ref{circuit}b, the total average power $ P_\textrm{avg}(\omega) = \operatorname{Re}\{ v_\textrm{d,rms}^2/Z_\textrm{tot} \}$ consumed by the MMT at $\omega_o$ is then evaluated through the expression
\begin{align}\label{pwr}
\nonumber    P_\textrm{avg}(\omega_o) &= \frac{v_\textrm{d,rms}^2}{\left(R_c + \frac{\Gamma_o^2}{b}\right)^2 + \left(\omega_o L_c\right)^2} \left(R_c + \frac{\Gamma_o^2}{b}\right) 
\\&= i_\textrm{d,rms}^2 \left(R_c + \frac{\Gamma_o^2}{b}\right)
\end{align}
where $i_\textrm{d,rms}$ represents the rms value of the current $i_\textrm{d}$ flowing through the drive coil. From Eq.\ref{pwr}, we see that $P_\textrm{avg}(\omega_o)$ is the sum of the power dissipated in the drive coil ($i_\textrm{d,rms}^2 \, R_c$) and the mechanical resonator ($i_\textrm{d,rms}^2 \, \Gamma_o^2/b$). When the damping $b$ is very small, we can approximate $P_\textrm{avg}(\omega_o) \approx v_\textrm{d,rms}^2 b / \Gamma_o^2$, implying that the power dissipated in the MMT reduces as the mechanical damping $b$ reduces.


\section{Experimental Results and Discussion}\label{Results_singlerotor}

In this section we present experimental demonstration of MMTs using both single-rotor and multi-rotor architectures. 

\subsection{Single-rotor MMT prototype and measurement setup}

Fig.~\ref{prototype} presents a single-rotor MMT prototype constructed using N52 grade NdFeB magnets (of square cross-section with $l_s$ = 12.7 mm and axial length $l_h$ = 50.8 mm) as the rotor and stators. The frame of the device is made of laser-cut acrylic sheets, and the rotor is mounted on the frame using aluminum end caps and stainless steel bearings. Stator magnets are inserted into separate acrylic holders and are fixed in position at a center-to-center distance of 16.5 mm from the rotor axis. A 170-turn air-core coil inductor made with AWG 18 enameled copper wire is used to drive the mechanical resonator. This drive coil is supplied with an ac voltage using a power amplifier (Pyle PTA1000) with sinusoidal input provided by a laboratory function generator (HP 33120A). An ac current transducer (Mastech MS3320) measures the current flowing through the coil. The average power consumed by the MMT is calculated by measuring the amplifier output voltage in conjunction with the coil current. A calibrated flux-gate magnetometer (Texas Instruments DRV425EVM) is used as the receiver to measure the total magnetic field vector component along the $\hat{x}$-direction at position $r_x \hat{x}$ as described previously in Fig.~\ref{device}b, with the help of a lock-in amplifier (Stanford Research Systems SR860). The received magnetic field can be analytically expressed as
\begin{linenomath}
\begin{align}
    {B}_\textrm{tot}(r_x,t) = -\frac{\mu_o M_r \sin(\theta)} {2 \, \pi r_x^3} + \frac{\mu_o N_c A_c}{2\pi (d_{cr}+r_x)^3} \, i_d
\end{align}
\end{linenomath}
where $\theta$ and $i_d$ are both functions of time. Using the distance dependence, and converting to the rms value, we can extrapolate the field at the receiver ${B}_\textrm{tot,rms}(1 \textrm{ km})$ for estimating performance at range. 

\subsection{Frequency response and signal modulation}
\label{results_singlerotor_freq}

\begin{figure}[!t]
     
    \centering
    \includegraphics[width=\linewidth]{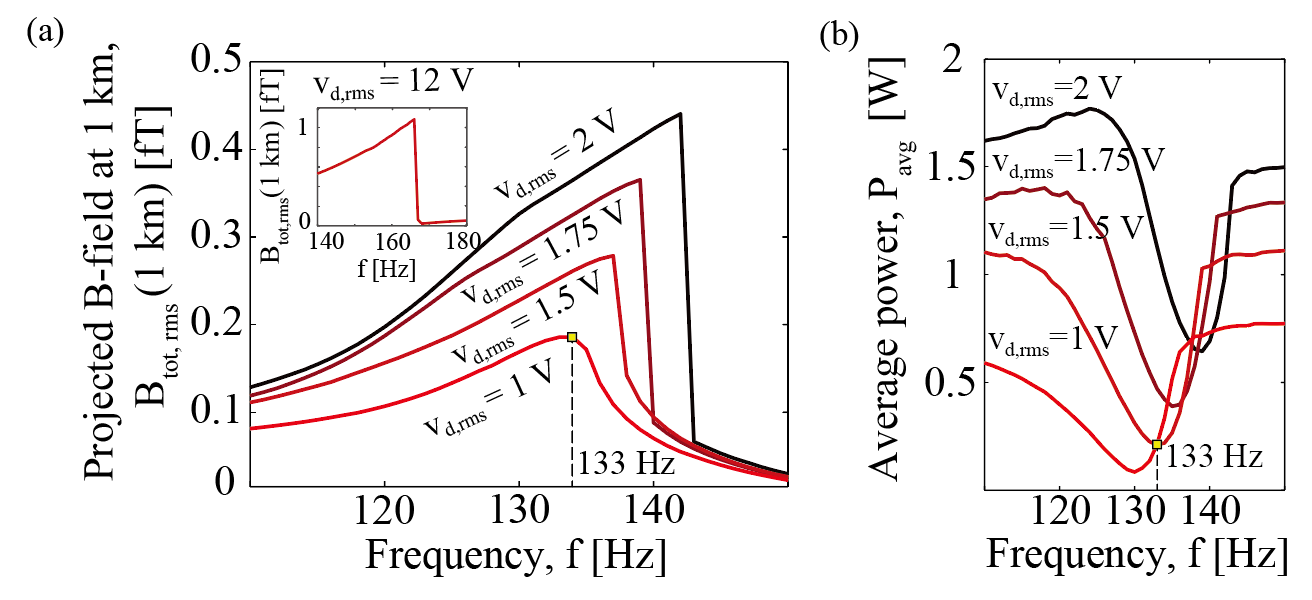}
    \caption{\textbf{(a)} Measured frequency response of magnetic field produced by our single-rotor MMT during an upward sweep of the drive frequency. The device exhibits a nonlinear stiffening behavior with increasing drive voltage, reaching ${B}_\textrm{tot,rms}(1\textrm{ km})$ = 1 fT for $v_\textrm{d,rms}$ = 12 V (inset)
    \textbf{(b)} Frequency response of the average power consumed by the MMT for the drive voltages in (a).
}\label{freqResp}
      
\end{figure}

We characterize the resonance frequency, magnetic field strength, and the corresponding power consumption of the prototype single-rotor MMT by studying its frequency response. Fig.~\ref{freqResp}a shows example amplitude response curves for the device at various drive voltages. Since the single-rotor MMT is predicted to exhibit a stiffening nonlinearity (Fig.~\ref{torquefig}), we perform upward-directed frequency sweeps in order to access the stable branch of the nonlinear response curve and to generate the highest magnetic field strength. 
   
As discussed in Eq.~\ref{mech_freq}, the magnetic field generated by the oscillating rotor at the receiver is maximized at the mechanical resonance frequency for a specific drive voltage. Since the effect of coil inductance and coupling is negligible as discussed in Section \ref{se:LEM}, we can approximate the frequency corresponding to the maximum magnetic field in Fig.~\ref{freqResp}a as the mechanical resonance frequency $f_\textrm{nonlin}=\omega_\textrm{nonlin}/2\pi$. We additionally see from Fig.~\ref{freqResp}a that this resonance increases with drive voltage ($v_\textrm{d,rms}$) as the nonlinear term specified in Eq.~\ref{backbone} becomes dominant at higher oscillation amplitude, thus experimentally verifying the nonlinear stiffening effect. In the nonlinear regime, this MMT has a mechanical resonance frequency of 166 Hz at the desired magnetic field ${B}_\textrm{tot,rms}(1 \textrm{ km}) = $1.09 fT (Fig.~\ref{freqResp}a inset), which is in close agreement with the theoretical estimate of 168 Hz predicted by the backbone curve in Eq.~\ref{mech_freq}. 

In Fig.~\ref{freqResp}b we present measurements of the average power consumption corresponding to the traces in Fig.~\ref{freqResp}a. For a fixed drive voltage, the power measurements show the predicted RLC tank behavior with a minimum occurring at the electrical resonance frequency (where $Z_\textrm{tot}$ in Eq.~\ref{Z} reaches its lowest magnitude), which is slightly lower than the frequency where the total magnetic field is maximized. We note that this single-rotor MMT prototype requires an average power of 21 W to generates ${B}_\textrm{tot,rms}(1 \textrm{ km}) = $ 1.09 fT. In the present design, the major sources of power dissipation in the mechanical resonator include frictional losses in the bearings, air damping, as well as magneto-electro-mechanical losses (eddy current losses, hysteresis losses, etc.) in the stators. The power consumption may be further reduced by decreasing the mechanical damping coefficient by using vacuum enclosures, laminated stators to mitigate eddy current losses, and low-loss bearings.

\begin{figure}[!t]
     
    \centering
	\includegraphics[width=0.75\linewidth]{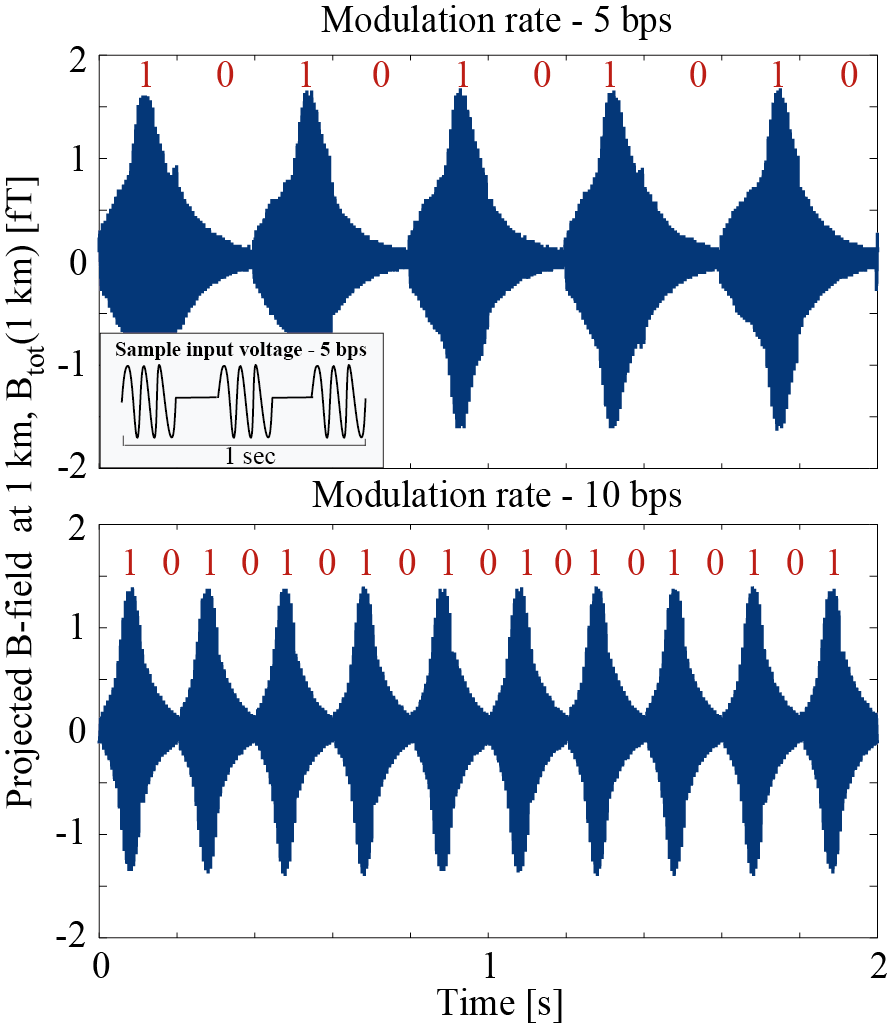}
	\caption{ Measured time-domain signals showing amplitude modulation of a carrier using On-Off Keying at 5 and 10 bps. Inset shows an illustration of the drive voltage encoded with a binary message signal at 5 bps modulation used to actuate the MMT. } \label{mod}
      
\end{figure}

In order to demonstrate transmission capability for short messages, we implemented On-Off Keying (OOK) \cite{OOK1,OOK2} as a simple amplitude modulation scheme. This is readily achieved by varying the drive voltage corresponding to the message bit (Fig.~\ref{mod} inset), with drive off representing a `0' bit and drive on representing a `1' bit. Modulation examples with ${B}_\textrm{tot,rms}(1 \textrm{ km})>1$ fT carrier magnetic field are presented in Fig.~\ref{mod}, showing OOK modulation at 5 and 10 bits per second. Compared to the envelope of the drive voltage signal, which is a square wave, the output magnetic field waveform envelope is shaped by the time constant $\lambda = 2I/b$ associated with mechanical damping in the resonator. Higher modulation rates can be achieved by reducing the time constant e.g., by reducing the moment of inertia $I$ of the MMT rotor(s).
\subsection{Determining the carrier frequency}

The mechanical resonance of an MMT determines the best carrier signal frequency for the transmitter. This resonance is in turn determined by the rotor moment of inertia as well as the torsional stiffness of the magnetic restoring spring, both of which are paths for frequency adjustment. 

\subsubsection{Tuning via torsional stiffness}

We previously found in Eq.~\ref{stiffness} that the torsional stiffness of the rotor magnetic spring can be controlled by changing the rotor-stator distance $d_\textrm{rs}$ \cite{Ali}. In Fig.~\ref{FreqCtrl}a we present experimental data for the measured dependence of mechanical resonance frequency $f_\textrm{meas}$ vs distance $d_\textrm{rs}$, for the single-rotor MMT. In each case we set the device to generate ${B}_\textrm{tot,rms}(1\textrm{ km}) \approx$ 1.1 fT so that we can make a fair comparison considering the amplitude dependence of $\omega_\textrm{nonlin}$ as discussed in Eq.~\ref{backbone}. The point dipole model predicts the relation $f_o \propto d_\textrm{rs}^{-1.5}$. However, the experimentally measured trend is more gradual at $f_\textrm{meas} \propto d_\textrm{rs}^{-1.09}$, likely due to the near-field effects discussed earlier. In practice, this empirical relation can serve as an aid to calculate the $d_\textrm{rs}$ required to obtain operation frequencies below 166 Hz with this MMT prototype. The minimum physical clearance requirement between rotor and stator sets a lower bound for $d_\textrm{rs}$, and limits the maximum frequency that can be achieved by a given set of rotor and stators.

\subsubsection{Tuning via rotor moment of inertia}

\begin{figure}[t]
     
    \centering
	\includegraphics[width=\linewidth, clip=true, trim=0in 0.4in 0in 0in]{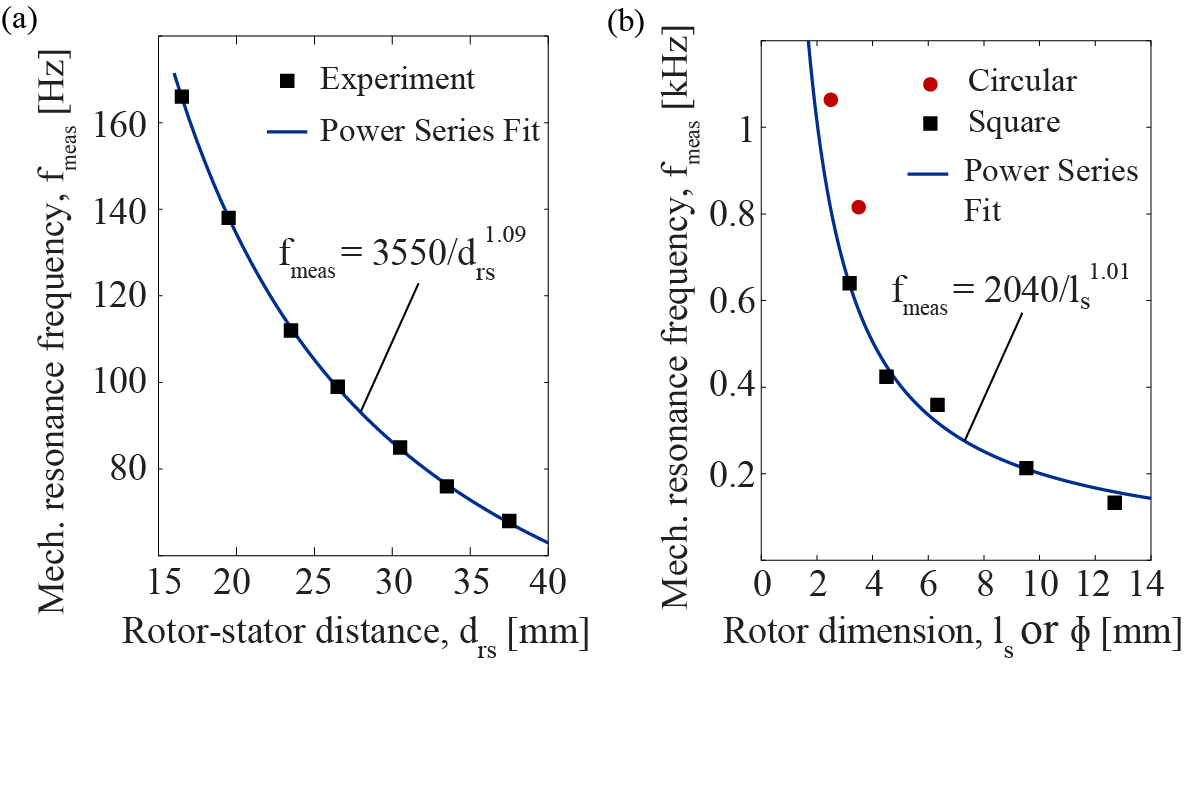}
    \caption{
    Variation of the mechanical resonance frequency vs rotor-stator distance and rotor shape. 
    \textbf{(a)} Measured mechanical resonance frequency of single-rotor MMT in Fig.~\ref{prototype} for various $d_\textrm{rs}$. 
    \textbf{(b)} Measured frequency for MMT variants having cuboidal and cylindrical rotors. Cuboidal rotors have square cross-section rotors indicated by their side length $l_s$. Circular cross-section rotors are indicated by the rotor diameter $\phi$. The power series relation is fitted only for the results of MMTs with square cross-section rotors.}\label{FreqCtrl}
      
\end{figure}

As discussed above, the mechanical resonance frequency can also be increased by reducing the moment of inertia of the rotor. We demonstrate this by replacing the rotor magnets to those with different cross-sectional area while keeping their axial length the same. Four variations of the single-rotor MMT (square cross-section $l_s$ = 9.5 mm, 6.3 mm, 4.5 mm and 3.1 mm) were assembled, with axial length fixed to $l_h$ = 50.8 mm, and with the same set of stators placed $d_\textrm{rs}$ = 16.5 mm away.

In Fig.~\ref{FreqCtrl}b we present the experimentally measured mechanical resonance frequency of these MMT prototypes. For an MMT using a square cross-section rotor and a specific set of stators (i.e., constant $M_s$), the mechanical resonance frequency trend can be estimated from Eq.~\ref{backbone} as $f_o \propto 1/l_s$ since both inertia $I$ and linear torsional stiffness $\kappa_1$ scale linearly with the rotor volume. The experimental results show good agreement to this inverse relationship and the fitted equation can be used as a guide to choosing an appropriate rotor size for the desired operational frequency. The empirical relationship predicts that 1 kHz may be achieved with a square cross-section rotor having a very small side length of $l_s= 2$ mm. However, manufacturing of high aspect ratio magnets with small square cross-section is challenging since NdFeB is a very brittle material. Alternatively, the moment of inertia can be further reduced by changing to a cylindrical rotor with the same cross-section area (albeit circular) and total volume as a square rotor. Such a cylindrical rotor has a diameter $\phi = 2l_s/\sqrt{\pi}$ and moment of inertia that is 4\% lower than an equivalent square cross-section rotor with side length $l_s$.  This allows MMTs with cylindrical rotors to achieve higher resonance frequency while not significantly compromising mechanical durability. Fig.~\ref{FreqCtrl}b additionally presents the mechanical resonance frequency of two MMT designs that use cylindrical rotors, with a resonance exceeding 1068 Hz demonstrated using a 2.5 mm diameter rotor. 

While decreasing the rotor moment of inertia helps achieve higher mechanical resonance frequency for the MMT, it reduces the total oscillating magnetic moment of the resonator. One simple path to compensate for this is to increase the length of the rotor for the same cross-section area, since the mechanical resonance frequency is independent of rotor length. This is because the rotor moment of inertia $I$ and linear stiffness $\kappa_1$ scale linearly with the rotor volume. However, due to the aforementioned brittleness of NdFeB this is not a practical option. As an alternative, we explore multi-rotor designs in the next section to address this trade-off.

\subsection{Scalability with Multi-rotor MMTs}

\subsubsection{Mode design}

To increase the magnetic moment of the MMT while maintaining a high frequency, one route is to distribute the total magnetic moment $M_r$ across N multiple rotors that individually have a smaller moment of inertia. For a system with N degrees of freedom, however, there exist N distinct eigenmodes. For the MMT application we are most interested in the mode that produces the largest time-varying magnetic field, i.e., the mode in which all the rotors undergo synchronized oscillatory motion.  

\begin{figure}[!t]
     
    \centering
	\includegraphics[width=0.8\linewidth]{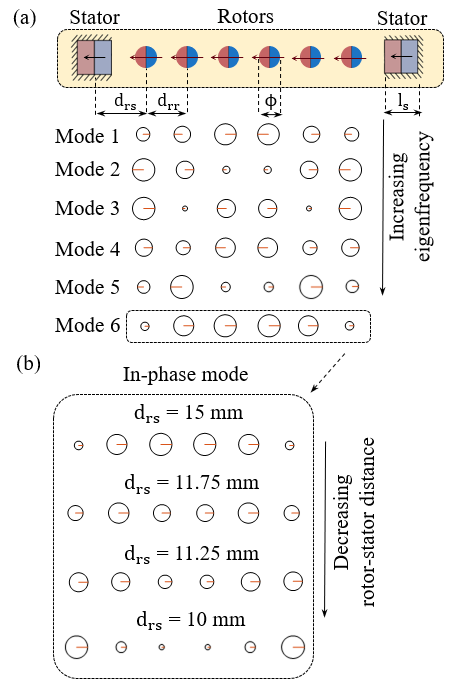}
    \caption{ Schematic and eigenmode analysis of a six-rotor mechanical resonator with cylindrical rotors showing rotor-rotor and rotor-stator separation by $d_\textrm{rr}$ and $d_\textrm{rs}$ respectively. \textbf{(a)} An example set of six eigenmodes is shown for a six-rotor MMT with $\phi =$ 3.5 mm rotor, $l_s$= 12.7 mm stators, $d_\textrm{rr}$ = 5.5 mm and $d_\textrm{rs}$ = 15 mm. All magnets are assumed to be N52 grade with $l_h$ = 76.2 mm.  The circle diameter is proportional to the normalized amplitude and the direction of straight line indicates the relative phase between the different rotors, with 0 phase pointing to the right and $\pi$ phase pointing to the left (arbitrary reference). \textbf{(b)} Evolution of the shape of the in-phase mode (Mode 6) with rotor-stator distance $d_\textrm{rs}$.
}\label{Multirotor_eigenmode}
      
\end{figure}

In order to understand the effective stiffness of this mode, we explore the case of a multi-rotor MMT with six identical cylindrical rotors and two cuboidal stators as shown in Fig.~\ref{Multirotor_eigenmode}a. Here, each rotor experiences a restoring torque from adjacent stators and rotors. While the rotor-stator interaction can be evaluated through Eq.~\ref{stiffness}, the rotor-rotor interaction for small oscillation amplitudes can be expressed (via dipole approximation) \cite{inbar} as 
\begin{equation}\label{torquerotor}
    {\tau}_{{r_i}{r_j}}= \frac{\mu_o {M_r}^2}{4 \pi {d}_{r_i r_j}^3}\left(2 \theta_i + \theta_j\right)
\end{equation}    
where ${\tau}_{r_i r_j}$ is the torque acting on rotor $i$ due to rotor $j$, $d_{r_i r_j}$ is the center-to-center distance between the rotors, $\theta_{i,j}$ are the respective oscillation amplitudes. Similar to the rotor-stator stiffness, the rotor-rotor stiffness is dictated by the rotor dipole moments, rotor-rotor separation and by their angular displacements. Now, using Eq.~\ref{torquerotor} and Eq.~\ref{stiffness} we can solve for the six eigenmodes of the example multi-rotor MMT by including the magnetic interaction from the first- and second-nearest neighbours of each rotor. The mode shapes are presented in Fig.~\ref{Multirotor_eigenmode}a. We observe that the mode in which all rotors undergo synchronized motion (labeled as Mode 6) has the highest frequency among all available eigenmodes. This is expected since, according to Eq.~\ref{torquerotor}, the restoring torque $\tau_{r_i r_j}$ is greatest when adjacent rotors have angular displacements $\theta_{i,j}$ in the same direction. Moreover, this mode produces the largest time-varying magnetic field at the receiver since the oscillatory motion of the rotor dipoles is in-phase.

A key question is whether there is an optimal mode shape by which we can produce the highest time-varying magnetic field for a given amount of kinetic energy in the mechanical resonator. Let us consider the simple case of a system with two rotors only, oscillating with rotational amplitudes $\theta_\textrm{avg}+\delta$ and $\theta_\textrm{avg}-\delta$ respectively, where $\theta_\textrm{avg}$ is the average of the oscillation amplitudes and $\delta<\theta_\textrm{avg}$. The total kinetic energy in this system can be expressed as $KE \propto 2\theta_\textrm{avg}^2 + 2\delta^2$ for a magnetic field ${{B}}_\textrm{mech,rms} \propto 2\theta_\textrm{avg}$ for small angular displacements (Eq.~\ref{bamp}). When the two rotors have similar oscillation amplitude (i.e., $\delta \rightarrow 0$), we obtain a lower total energy requirement for the same time-varying field. This argument can be readily extended to a multi-rotor system, where the best efficiency is obtained when all the rotors have approximately equal amplitude of oscillation.

The oscillation amplitude of the rotor depends on the local stiffness within the multi-rotor system. In the case of a hypothetical system with infinite rotors, all the rotors would experience the same local stiffness and hence oscillate with the same amplitude. However, in a finite system, the local torsional stiffness of the rotors towards the ends of a linear array is generally different than for rotors in the center. This leads to a boundary effect causing the end rotors to have a smaller oscillating amplitude (e.g., Mode 6 in Fig.~\ref{Multirotor_eigenmode}a). To achieve a more uniform mode shape, we can compensate for the non-uniformity in local stiffness by varying the rotor-stator distance, as shown by simulations in Fig.~\ref{Multirotor_eigenmode}b. At some optimum $d_\textrm{rs}$, all the rotors have approximately the same local stiffness and oscillate with equal amplitude. Unfortunately, it is challenging to accurately calculate this distance using simple analytical models due to the nonlinear and near-field effects discussed earlier. However, one can make ad hoc changes to $d_\textrm{rs}$ in experimental multi-rotor prototypes to achieve the optimum mode shape. 

\subsubsection{Modular approach and experimental testing}

\begin{figure*}[ht]
     
    \centering
	\includegraphics[width=0.8\textwidth]{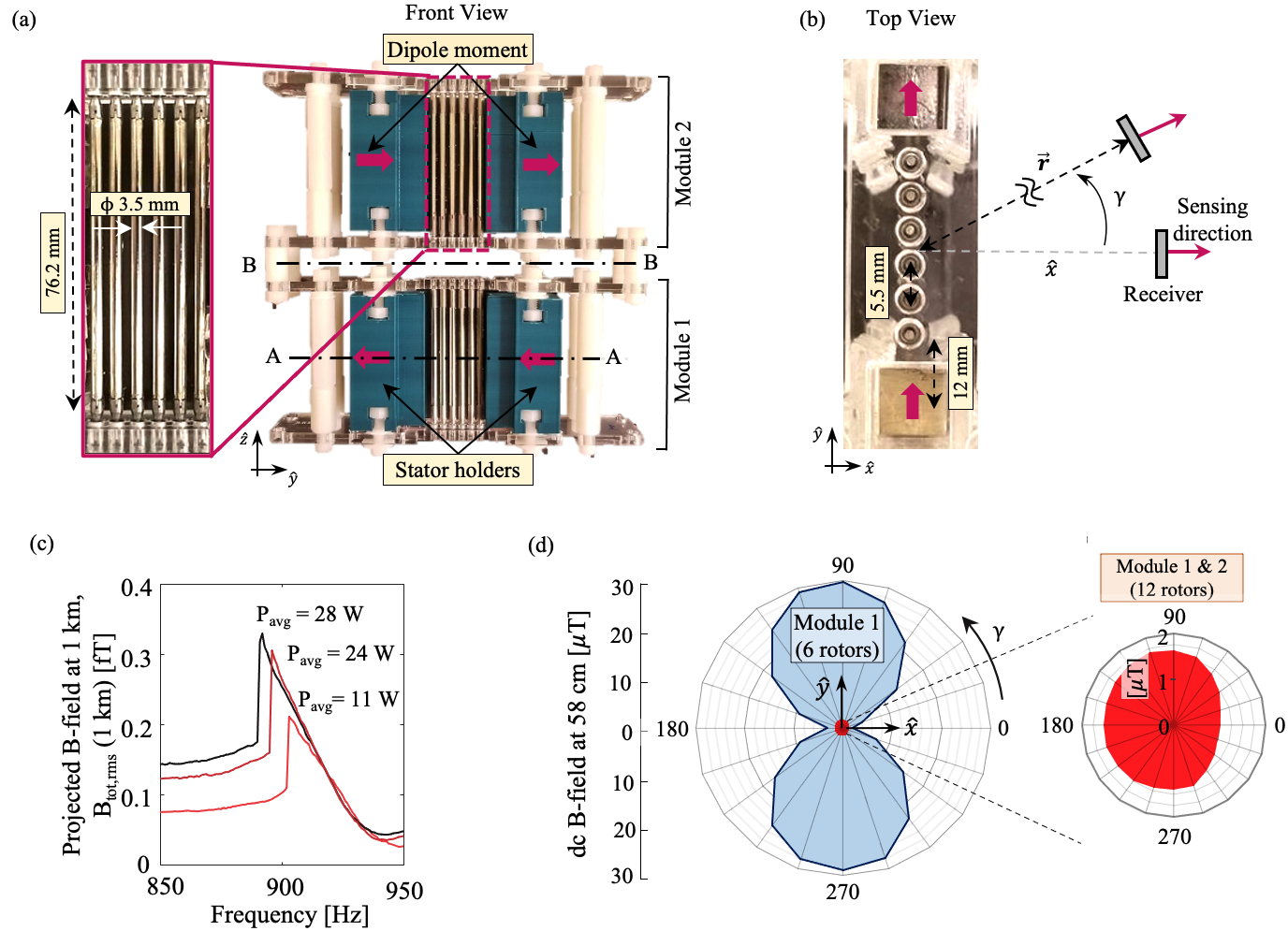}
    \caption{ \textbf{(a)} Photograph of a multi-rotor MMT using two modules. Each module has 6 cylindrical rotors ($\phi =$ 3.5 mm and $l_h$ = 76.2 mm ) and 2 stators (square cross-section with $l_s$ = 12.7 mm and $l_h$ = 76.2 mm). Line A-A shows the mid-plane of module 1 and line B-B shows the mid-plane of the full multi-rotor MMT.
    \textbf{(b)} Top view showing the relative position of the MMT and the receiver (flux-gate sensor). $\gamma$ denotes the angular position of the receiver with respect to $\hat{x}$.
    \textbf{(c)} Frequency response of magnetic field produced by our multi-rotor MMT during downward sweep of the drive frequency measured at $\gamma$ = 0. The device shows softening nonlinearity with increasing drive voltage, indicated by proportional increase in $P_\textrm{avg}$ at the peak magnetic field. 
    \textbf{(d)} Radially directed dc magnetic field around the 6-rotor MMT (module 1), measured in section A-A for various receiver angular positions $\gamma$. A similar plot for the 12-rotor MMT (modules 1 and 2) measured in section B-B shows a significantly lower peak dc magnetic field due to the opposing dipole orientation of the two modules as shown in (a). 
    }\label{Multirotor}
      
\end{figure*}

The multi-rotor approach also facilitates the development of identical MMT modules that operate together for the generation of larger carrier magnetic fields. For instance, two 6-rotor modules when stacked appropriately (Fig.~\ref{Multirotor}a) can function as a larger MMT system that can produce double the time-varying field. The modules in this example are driven using separate coils that are connected in series to a common voltage source. This drive configuration ensures equal current flow through both drive coils resulting in the same magnitude of drive torque for each module and uniform mechanical excitation.

Experimental results for the example system with two modules are presented in Fig.~\ref{Multirotor}c. We see that the in-phase mode for this MMT is around 900 Hz and its frequency decreases with increasing input power. This result agrees with a reduction of the stiffening nonlinearity that is predicted by our numerical simulations of rotor magnets having small cross-section area (Fig.~\ref{torquefig}). Since the overall architecture of the multi-rotor device is similar to a single-rotor MMT, amplitude modulation and frequency control techniques discussed earlier can be adopted here as well. In the multi-rotor MMT, the rotor-rotor distance offers an additional degree of freedom to vary the operational frequency. 

At rest, the magnetic dipole moment of all the rotors and stators in a module self-align in the same direction and act as a very large net magnetic dipole, which can make storage and handling rather challenging due to the large dc field. Fortunately, an additional feature of the modular approach is the possibility of constructing an MMT using multiple modules with opposing dipole orientations that cancel the static dc magnetic field. Such a reduction in the dc field can significantly simplify storage when the MMT is not in use. Even when these static dipoles cancel each other (e.g., in the configuration of Fig.~\ref{Multirotor}a), the projection of their frequency-matched dynamic motion still adds constructively on the $\hat{x}$ axis, implying that the time-varying carrier field will not be sacrificed. 

To illustrate this feature, we can measure the radially directed dc magnetic field around an MMT as a function of the receiver's angular position $\gamma$ relative to the $\hat{x}$ axis, as illustrated in Fig.~\ref{Multirotor}b. In Fig.~\ref{Multirotor}d, we show that the radially directed dc magnetic field produced by a single module (Module 1) exhibits a two-lobe field pattern. This is expected since the radial dc magnetic field is maximum along the net dipole direction (i.e., along the $\hat{y}$ axis at $\gamma = 90^\circ$ and $270^\circ$) and diminishes to zero when measured perpendicular to the dipole (i.e., $\gamma = 0^\circ$ and $180^\circ$). The single-module MMT measurements shown here are taken on the X-Y mirror plane that cuts through the middle of the rotors (section A-A as indicated in Fig.~\ref{Multirotor}a). We now add a second module to this system with net dipole moment oriented in the opposite direction (Fig.~\ref{Multirotor}a) to produce a 12-rotor MMT. When measured on the X-Y mirror plane that cuts through the middle of this new system (section B-B as indicated in Fig.~\ref{Multirotor}b), the opposing dipoles cancel and the dc field drops to a very small value as shown in Fig.~\ref{Multirotor}d. 

\section{Conclusions}

In this work we have explored the design space for resonant magneto-mechanical transmitters, that can produce detectable ULF magnetic fields with frequencies ranging from sub-100 Hz to above 1 kHz. Scaling the frequency to the kHz range is achieved by using multi-rotor devices where the in-phase mode can readily approach kHz. Additionally, we show how a modular design approach can permit scaling of these systems to produce much higher magnetic fields while maintaining the operational frequency and canceling out the dc ambient field.

Since these systems have very low coupling to the earth-ionosphere waveguide, they cannot produce propagating ULF elctromagnetic waves such as those used for submarine communications over long distances \cite{Russia-ant,Us-ant2}. Even so, the fields produced by MMTs are highly detectable using compact magnetometers \cite{magnetometer1,magnetometer2} that are placed in the near-field environment. As a result, these systems are especially good at filling short-range radio communications dark zones, e.g., in conductive or radio-obstructed environments, by enabling low data rate communications such as text messaging. Some key applications in this space include emergency messaging and diver coordination in underwater environments, and below-earth communications in mines, tunnels, and collapsed buildings. Moreover, a constellation of fixed MMTs may also be usable as positioning beacons for underground positioning and navigation.

\section*{Acknowledgments}
This work was sponsored by the Defense Advanced Research Projects Agency (DARPA) grant HR0011-17-2-0057 under the AMEBA program.

\bibliographystyle{ieeetr}
\bibliography{main}

\end{document}